\documentclass[fleqn,usenatbib]{mnras}

\usepackage{newtxtext,newtxmath}
\usepackage[T1]{fontenc}
\DeclareRobustCommand{\VAN}[3]{#2}
\let\VANthebibliography\thebibliography
\def\thebibliography{\DeclareRobustCommand{\VAN}[3]{##3}\VANthebibliography}

\usepackage{graphicx}
\usepackage{amsmath}

\title[Nested Sampling for ARIMA Model Selection]{Nested Sampling for ARIMA Model Selection in Astronomical Time-Series Analysis}

\author[A. Naik and W. Handley]{Ajinkya Naik$^{1}$\thanks{E-mail: najinkya1313@gmail.com},
Will Handley$^{1}$\thanks{E-mail: wh260@cam.ac.uk} \\
$^{1}$Institute of Astronomy, University of Cambridge, Cambridge, CB3 0HA, UK\\
}

\date{}
\pubyear{\the\year{}}

\begin{document}

\label{firstpage}
\pagerange{\pageref{firstpage}--\pageref{lastpage}}
\maketitle

\begin{abstract}
The era of large-scale, high-cadence astronomical surveys demands efficient and robust methods for time-series analysis. ARIMA models provide a versatile parametric description of stochastic variability in this context. However, their practical use is limited by the challenge of selecting optimal model orders while avoiding overfitting. We present a novel solution to this problem by combining Autoregressive Integrated Moving Average (ARIMA) models with the Nested Sampling algorithm. Our method yields Bayesian evidences for model comparison and also incorporates an intrinsic Occam’s penalty for unnecessary model complexity. Using \texttt{JAX} and \texttt{Blackjax}, a vectorized ARIMA–Nested Sampling framework with GPU-acceleration support is implemented, allowing us to perform model selection across grids of Autoregressive (AR) and Moving Average (MA) orders, with efficient inference of selected model parameters. We validate the approach using simulated time series with known ground-truth parameters and demonstrate accurate recovery of both model order and parameters. We then apply the method to several astronomical datasets, including the historical sunspot number record, stellar light curves of KIC 12008916 and Kepler 17 from the Kepler mission, and quasar light curves of 3C 273 and S4 0954+65 from the TESS mission. For all cases, except Kepler 17, the ARIMA models selected by this method were able to accurately model the stochastic variability in the time series data as well as produce accurate multi-step ahead forecasts for the sunspot number time series. Our results demonstrate that nested sampling offers a rigorous and computationally tractable alternative to autoregressive model selection in astronomical time-series analysis.
\end{abstract}


\section{Introduction}
Time-series analysis is a crucial tool in the field of time domain astronomy. As next-generation astronomical surveys and telescopes are set to yield unprecedented volumes of complex time series data, a robust and scalable method for their analysis is necessary. Some examples of such methods are frequency-domain analysis \citep{Lomb1976,Scargle1982}, Gaussian Processes \citep{ForemanMackey2017} and Machine Learning methods \citep{Richards2011,baron2019machinelearningastronomypractical}. Within this diverse landscape of methodologies, parametric autoregressive modelling offers a complementary approach to time series analysis in the form of ARIMA models \citep{discretearimaastronomy,AKHTER2020100403,asteroidlightcurve}.
\par
The ARIMA (Autoregressive-Integrated-Moving Average) framework is a suite of models employed in analysing and forecasting time-series data from various domains including, but not limited to, economics, finance and climate science. The Autoregressive (AR) component of ARIMA was first introduced by \citet{Yule1927} to study the number of sunspots. These models capture the autocorrelation in a time series by linearly regressing the present value on its own past (lagged) values. Moving Average (MA) models on the other hand, operate on a similar principle by expressing the present value as a linear combination of past forecast errors or residuals. The idea of systematically combining these modelling methods can be traced back to \citet{box1976time}, who formalized them into ARIMA with the introduction of the Integrated (I) component to handle non-stationary time series. These are labelled as ARIMA$(p,d,q)$ models, characterized by the respective orders $p$, $d$ and $q$ of the Autoregressive (AR), Integrated (I) and Moving Average (MA) components. 
\par
ARIMA models are seldom used in the analysis of astronomical time-series data. The primary reason is that they require evenly sampled data in time, which is often in contrast to ground-based astronomical datasets due to their irregular cadences. Nevertheless, ARIMA modelling can still be applied, with reasonable success, by binning light curve data with moderately irregular cadences \citep{10.3389/fphy.2018.00080}. Upcoming facilities such as the Vera C. Rubin Observatory’s Legacy Survey of Space and Time (LSST) \citep{2019ApJ...873..111I}, the Roman Space Telescope \citep{Spergel2015,Johnson2020}, and ongoing space missions like TESS \citep{Ricker2015} and Gaia \citep{Gaia2016} will also produce adequately regular and high-cadence time-series data suitable for ARIMA modelling. 

The generality and flexibility of the ARIMA framework allows it to efficiently model a diverse range of astronomical time series. However, this very flexibility comes with the risk of over-parametrization and overfitting. It is non-trivial to choose the optimal $(p,d,q)$ order of the ARIMA model. The heuristic diagnostic approach taken by the standard Box-Jenkins methodology, commonly used for ARIMA model specification and validation, may limit robustness in some cases \citep{markidakis_etal,MELARD2000497}. Model selection on the basis of the Akaike Information Criterion (AIC) \citep{Akaike1974} and Bayesian Information Criterion (BIC) \citep{Schwarz1978} relies on maximum likelihood optimization, which may bias the model selection, especially for complex likelihood shapes \citep{Trotta01032008}.  
\par
The Nested Sampling algorithm used in Bayesian computation has shown promise in model selection problems in the context of astrophysics \citep{Skilling2006,ns_forphysicists,10.1214/23-SS144}. Recent advances in GPU-accelerated nested sampling \citep{yallup2026nestedslicesamplingvectorized} have reduced the computational costs associated with this method, allowing it to be implemented for a variety of model selection problems \citep{ormondroyd2025dynamicsystematicbayesianmodel,lovick2025highdimensionalbayesianmodelcomparison,leeney2025bayesiananomalydetectionia,prathaban2025gravitationalwaveinferencegpuspeed,yallup2025particlemontecarlomethods}. While the use of Bayesian inference for autoregressive models  has been explored in the econometrics literature \citep{BARNETT1996237,Marriott1996BayesianAO}, such approaches have relied on MCMC sampling methods. Using nested sampling in conjunction with ARIMA models can thus provide an alternative solution to the problem of selecting the optimal $(p,d,q)$ order, with the added benefit of returning posterior samples for the ARIMA parameters. The previously discussed risk of overfitting is avoided with an in-built Occam's penalty imposed on the Bayesian evidences of over-parametrized model fits. The main contribution of our work is \emph{NestAR} -- a vectorized ARIMA-Nested Sampling framework with GPU-acceleration support which can be applied for modelling diverse astronomical time-series datasets.
\par
The structure of this paper will be as follows -- in the following section, we will present a review of the mathematical formalism of ARIMA models, Bayesian inference for model selection, and a brief overview of the nested sampling algorithm. Section \ref{methodology} then describes the methodology adopted, details of using the nested sampling algorithm for ARIMA model selection, and defines the criterion we used for validating the model fit. The results of applying this framework on artificial and real astronomical time-series data are presented in Section \ref{results}. We discuss potential future directions for this work and end with conclusions in Section \ref{conclusions}.
\section{Background}
Subsection \ref{arima} outlines the mathematical background and properties of ARIMA models. For a more comprehensive exposition of ARIMA models, readers are directed to the standard text of \citet{box1976time}. In subsection \ref{bayesian}, we introduce Bayesian inference, its associated terminologies, and discuss the role of the Bayesian evidence in model selection. 
\subsection{ARIMA Framework}
\label{arima}
The general forecasting equation for ARIMA modelling is a combination of the Autoregressive (AR) and Moving Average (MA) model equations. An AR($p$) equation to obtain the forecast $\hat{y}_t$ for an observed point $y_t$ of the time series is
\begin{equation}\label{areqn}
    \hat{y}^{\textrm{AR}}_t = c + \sum_{a=1}^p\phi_a y_{t-a} + \epsilon_t
\end{equation}
It involves a linear weighted sum over $p$ lagged values of the time series with a constant intercept term $c$ (usually related to the long-term mean or "drift" of the time series) and a random forecasting error $\epsilon_t$ added to the forecast. The random noise $\epsilon_t$ is assumed to be homoscedastic and drawn from a normal distribution. Similarly, an MA($q$) model is expressed by a linear weighted sum of $q$ lagged forecast errors, again with a constant term $c$ and a random error $\epsilon_t$ 
\begin{equation}\label{maeqn}
    \hat{y}^{\textrm{MA}}_t = c + \sum_{m=1}^q\theta_m\epsilon_{t-m} + \epsilon_t
\end{equation}
Equations \ref{areqn} and \ref{maeqn} together represent an ARMA process:
\begin{equation}\label{ARIMA}
    \hat{y}_t = c + \sum_{a=1}^p\phi_a y_{t-a} + \sum_{m=1}^q\theta_m \epsilon_{t-m} + \epsilon_t
\end{equation}
An ARMA process fundamentally assumes the time series data to be stationary. A time series is defined to be stationary if its statistical properties, such as mean, variance and autocorrelation, remain constant over time. The Integrated (I) component of ARIMA models is an additional tool for dealing with non-stationary time series using finite differencing. Its associated order $d$ is the number of times the raw observations are differenced to render the series stationary. After fitting the differenced time series to an ARMA process, the forecast values $\hat{y}_t$ are integrated back to recover the original sequence trend. 
\par
To facilitate a compact representation of the general ARIMA$(p,d,q)$ equation, the Backshift (or Lag) operator $\mathcal{B}$, with the action of shifting a quantity back by one time-step, is introduced: 
\begin{align}
    \mathcal{B} y_t  = y_{t-1} && \mathcal{B}\epsilon_{t} = \epsilon_{t-1}
\end{align}
The AR and MA components of the ARIMA process can then be written as
\begin{align}
   \label{arbackshift} (1-\phi_1\mathcal{B} -\ldots-\phi_p\mathcal{B}^p)y_t = \Phi^p(\mathcal{B})y_t \\ (1+\theta_1\mathcal{B} + \ldots+\theta_q\mathcal{B}^q)\epsilon_t = \Theta^q(\mathcal{B})\epsilon_t\label{mabackshift}
\end{align}
Here, $\Phi^p(\mathcal{B})$ and $\Theta^q(\mathcal{B})$ are the characteristic polynomials of the backshift operator for the AR and MA components. The differenced time series $Y_t$ associated with the Integrated (I) part can also be represented in terms of $\mathcal{B}$ as 
\begin{equation}\label{differenced}
(1-\mathcal{B})^d y_t
\end{equation}
Finally, combining Equations \ref{arbackshift}, \ref{mabackshift} and \ref{differenced} condenses the forecasting equation for an ARIMA$(p,d,q)$ model: 
\begin{equation}\label{ARIMAeqn}
    \Phi^p(\mathcal{B})(1-\mathcal{B})^dy_t = \Theta^q(\mathcal{B})\epsilon_t + c 
\end{equation}
\subsubsection{Stationarity and Invertibility Constraints}\label{stationarity_invertibility}
Even though the problem of non-stationarity in the time series data $y_t$ is solved by differencing, the ARIMA model fitted to the data must itself produce a stationary and stable time series forecast. This requirement manifests in the form of two mathematical constraints on the AR and MA weights $\phi_a$ and $\theta_m$ - the stationarity and invertibility constraints, respectively. Mathematically, these constraints are defined by imposing the following condition: all roots of the AR and MA characteristic polynomials - $\Phi^p(\mathcal{B})$ and $\Theta^q(\mathcal{B})$, must lie outside the unit circle. Therefore an ARIMA$(p,d,q)$ process is stationary and invertible if the absolute value of all the roots of $\Phi^p(\mathcal{B})$ and $\Theta^q(\mathcal{B})$ is greater than one. We will now mathematically motivate these conditions for an AR and MA process, respectively. 
\par
The stationarity constraint on the AR weights ensures that the modelled time series exhibits a stable and mean-reverting behaviour. It allows for the influence of past values to eventually decay with time, preventing the series from diverging to infinity. This can be demonstrated for a simple AR(1) model, for which the characteristic polynomial is
\begin{equation}
    \Phi^1(\mathcal{B}) = (1-\mathcal{B}\phi_1)
\end{equation}
This polynomial has the root: 
\begin{equation}
    \mathcal{B} = 1/\phi_1
\end{equation}
Imposing the stationarity constraint on this root, we get:
\begin{equation}\label{ar_1_stationarity}
|\phi_1|<1    
\end{equation}
The $t^{\text{th}}$ observation of an AR$(1)$ time series can be expressed indefinitely in terms of the lagged values (suppressing the constant term $c$ for simplicity):
\begin{equation}
    y_t = \phi_1y_{t-1} + \epsilon_t = \phi_1^2y_{t-2} + \phi_1\epsilon_{t-1} +\epsilon_t = \ldots
\end{equation}
Here, the ellipsis denote expressions written in terms of subsequently higher lags. Evidently, if Condition \ref{ar_1_stationarity} is violated, the influence of past values on the present value $y_t$ will grow without bound, resulting in a divergent, non-stationary time series. For a general, higher order AR$(p)$ process, the constraints on each individual weight $\phi_p$ become coupled and non-trivial. Thus, the unit-root stationarity condition ensures a stable and mean-reverting behavior of the time series. 
\par
The invertibility constraint is the dual to the stationarity condition. To see this, first note that any Moving Average process can be represented in terms of an infinite AR$(\infty)$ process \citep[Sec.3.1]{brockwell2009time}. The invertibility condition simply ensures that the ARIMA process admits a convergent infinite representation and subsequently, the present forecast errors $\epsilon_t$ can be expressed uniquely as a linear combination of past observations. We demonstrate this again for a simple MA$(1)$ model and extend it to the general case of an MA$(q)$ process. For MA$(1)$, the condition on the roots of the characteristic polynomial $\Theta^1(\mathcal{B})$ results in the condition on the MA weight:
\begin{equation}\label{invertibility}
    |\theta_1|<1
\end{equation}
Rewriting the forecasting equation for an MA$(1)$ model in terms of $\epsilon_t$:
\begin{equation}
    \epsilon_t = y_t - \theta_1\epsilon_{t-1} = y_t + \theta_1^2\epsilon_{t-2} - \theta_1y_{t-1} = \ldots
\end{equation}
Continuing indefinitely, we get the infinite AR$(\infty)$ representation of an MA$(1)$ process:
\begin{equation}
 \epsilon_t = y_t - \theta_1y_{t-1} - \theta_1^2y_{t-2} - \theta_1^3y_{t-3}\ldots
\end{equation}
The above series is convergent only when Condition \ref{invertibility} is satisfied. Moreover, the autocorrelation at lag $1$ is given by:
\begin{equation}
    \rho_1 = \frac{\theta_1}{1+\theta_1^2}
\end{equation}
There is a degeneracy in this autocorrelation for $\theta_1$ and $1/\theta_1$. Therefore, the constraint $|\theta_1|<1$ also enforces a unique representation of the MA$(1)$ model by its parameters. This requirement of a unique model representation extends to higher orders. For an MA$(q)$ process, there exists a $2^q$-fold degeneracy where $2^q$ distinct sets of parameters yield identical autocorrelation structures. This occurs because any root $z_i$ of the MA characteristic polynomial $\Theta^q(\mathcal{B})$ can be replaced by its reciprocal $1/z_i$ without altering the autocorrelation. This degeneracy is removed by again ensuring that each of these $q$ roots must lie outside the unit circle ($|z_i|>1$).
\subsection{Bayesian Inference}\label{bayesian}
Given a dataset $D$, and a model $\mathcal{M}$ characterized by a vector of parameters $\boldsymbol{\theta}$\footnote{We use a boldface notation to distinguish the full parameter vector $\boldsymbol{\theta}$ from the MA coefficient $\theta_m$}, Bayesian inference updates the prior distribution on the model parameters $P(\boldsymbol{\theta}|\mathcal{M})$ to the posterior distribution $P(\boldsymbol{\theta}|D;\mathcal{M})$ via Bayes' theorem: 
\begin{equation}
    P(\boldsymbol{\theta}|D;\mathcal{M}) = \frac{P(D|\boldsymbol{\theta};\mathcal{M}) P(\boldsymbol{\theta}|\mathcal{M})}{P(D|\mathcal{M)}} = \frac{\mathcal{L}(D|\boldsymbol{\theta})\pi(\boldsymbol{\theta})}{Z}
\end{equation}
Here $\mathcal{L}(D|\boldsymbol{\theta})$ is the Likelihood function and the normalization constant $Z$ is the Bayesian evidence:
\begin{equation}\label{evidence}
    Z = \int \mathcal{L}(D|\boldsymbol{\theta}) \pi(\boldsymbol{\theta}) \hspace{4px} d\boldsymbol{\theta}
\end{equation}

The posterior distribution is the primary object of interest in parameter estimation. From it, one can extract point estimates, credible intervals, and correlations between parameters. A natural measure of the information gained by moving from prior to posterior, that is, of how much the data have constrained the parameter space is the Kullback–Leibler (KL) divergence \citep{1320776d-9e76-337e-a755-73010b6e4b64}:
\begin{equation}
    D_\mathrm{KL} = \int P(\boldsymbol{\theta}|D) \log{\frac{P(\boldsymbol{\theta}|D)}{\pi(\boldsymbol{\theta})}} \hspace{4px}d\boldsymbol{\theta}
\end{equation}
A large $D_\mathrm{KL}$ indicates that the data are highly informative about the parameters; a value near zero indicates that the posterior closely resembles the prior and the data carry little constraining power.
\subsubsection{Model Comparison and Nested Sampling}
Beyond parameter estimation, Bayesian inference provides a natural framework for model comparison and selection through the Bayesian evidence \citep{Jeffreys1961, Kass1995}. For a set of competing models $\mathcal{M}_i$, the evidence $Z_i$ can also be interpreted as the conditional probability: $P(D|\mathcal{M}_i)$. Using Bayes' theorem for models, the posterior probability of a model $P(\mathcal{M}_i|D)$ can be computed:
\begin{equation}
    P(\mathcal{M}_i|D) = \frac{P(D|\mathcal{M}_i).P(\mathcal{M}_i)}{\sum_iP(D|\mathcal{M}_i).P(\mathcal{M}_i)}
\end{equation}

Assuming a uniform prior, $P(\mathcal{M}_i)=1/n$ across all $n$ competing models, the above equation reduces to:
\begin{equation} \label{modelposterior}
    P(\mathcal{M}_i|D) =\frac{P(D|\mathcal{M}_i)}{\sum_iP(D|\mathcal{M}_i)}= \frac{Z_i}{\sum_i Z_i}
\end{equation}

The evidence $Z_i$ associated with any model is thus a principled measure of overall model adequacy. This in turn is governed by the likelihood function $\mathcal{L}(D|\boldsymbol{\theta})$ as well as the size of the prior space over which the fit is performed. The latter factor ultimately acts as an Occam's razor that penalizes the evidence of an overfitting model. 

It is generally intractable to analytically evaluate the evidence integral (Equation \ref{evidence}), especially in higher-dimensional spaces. Nested Sampling is a widely-used tool which numerically performs both parameter estimation and model comparison \citep{ns_forphysicists}. It iteratively evaluates the evidence integral by exploiting the statistical properties of a set of compressing ``live points" in the parameter space. At each iteration, the lowest-likelihood live point is removed and replaced by a new sample drawn from the prior subject to a higher likelihood constraint. In vectorized implementations such as ours, multiple live points may be deleted and replaced simultaneously, enabling efficient parallelized exploration of the parameter space.
\section{Methodology}\label{methodology}
This section outlines the methodology adopted to apply the nested sampling algorithm for ARIMA models. We discuss details of the likelihood and prior functions in subsection \ref{subsec:likelihood_priors}, the nested sampler in subsection \ref{sampler_config}, and finally the model selection and validation procedures in subsections \ref{ARIMA_model_selection} and \ref{model_validation}, respectively. 
\par
The following quantities are treated as the set of parameters characterizing any given ARIMA model: the respective AR and MA weights $\phi_{a}$ and $\theta_{m}$, the missing initial data $D_0$, the standard deviation $\sigma$ associated with the present error $\epsilon_t$ and the unconditional mean of the time series $\mu$, which can be shown to be related to the intercept term $c$ through: 
\begin{equation}
    c = \mu (1- \sum_a^p\phi_a)
\end{equation}
\subsection{Likelihood and Priors}\label{subsec:likelihood_priors}
A Gaussian likelihood function is chosen for the $n$ observations of the given time series $D_t$ :  
\begin{equation}
    \mathcal{L}(D|\boldsymbol{\theta}) =  \prod_{t=1}^n \frac{1}{\sqrt{2\pi\sigma^2}}\exp{\left(\frac{-(D_t-\mathbf{\hat{y}_t})^2}{2\sigma^2}\right)}
\end{equation}
After fitting the ARIMA model, the resulting residuals are assumed to be independent Gaussian random variables with mean zero and constant variance ($\sigma^2$).
Since the noise model enters explicitly through the likelihood function, the ARIMA forecast values $\mathbf{\hat{y}_t}$ used in computing the likelihood should be completely deterministic. These deterministic forecasts $\mathbf{\hat{y}_t}$ are obtained from Equation \ref{ARIMAeqn} by setting $\epsilon_t=0$. For simplicity, we treat the missing initial values for the time series data $D_0$ as nuisance parameters in our inference. The initial forecast errors $\epsilon_0$ are set to zero, thus evaluating a likelihood which is conditional on $\epsilon_0=0$. 
\par
We now discuss the choice of prior distributions for the ARIMA model parameters. For $\sigma$, a reasonable choice of prior is a Half-Normal or a Half-Cauchy distribution with scale $\sigma'$. We adopt a relatively weak prior for $\sigma$ by choosing a half-normal distribution truncated from below at zero and scaled by $\sigma'=50$ in the native units (counts, normalized flux, etc.) of the analysed time series. We verified that the resulting evidences are insensitive to moderate variations in this scale parameter. The prior choice for the unconditional mean $\mu$ depends on the time series data under consideration. For differenced or de-trended time series, a normal prior centred at zero with unit scale in the native units of the time series is adopted for $\mu$. In other cases, a wide normal distribution centred around the expected long-term mean $\mu_0$ of the time series is used. The prior distribution for the $p$ missing initial values of $D_0$ is set equivalent to the prior for $\mu$, as we expect these values to lie close to $\mu$ for a stationary series. In the case of an ARIMA($1,d,1$) model, a uniform distribution bounded between -1 and 1 is the most straightforward choice of priors for the coefficients $\phi_a$ and $\theta_m$. The bounds are in accordance with the stationarity and invertibility conditions discussed in subsection \ref{stationarity_invertibility}. However, for higher-order ARIMA models, the constraints on the individual weights are coupled and hence the choice of prior distributions for them is non-trivial. 
\par
For this paper, we assign a mathematically constrained normal prior distribution $\mathcal{N}(0,\sigma'')$ to the ARMA weights and implement a ``rejection sampling" approach. The scale of the distribution $\sigma''$ is typically set to unity as we do not expect the time series under consideration to be very strongly auto-correlated. A sufficiently large set of $n_{\text{live}}$ points is sampled from this distribution. The points are tested for stationarity and invertibility by calculating the roots of the AR and MA characteristic polynomials, and imposing the conditions discussed in subsection \ref{stationarity_invertibility}. The parameter points passing this test are then added to a pool of valid particles. This process is continued iteratively until the desired number of valid live particles $n_{\text{live}}$ is reached in this pool. The normal prior distributions on the weights are also constrained to the stationary and invertible regions of the parameter space.
\par
Table~\ref{tab:prior_table} summarizes the prior distributions assigned to each parameter of the ARIMA model.

\begin{table}
\centering
\caption{Prior distributions for ARIMA model parameters.}
\label{tab:prior_table}
\begin{tabular}{ll}
\hline\hline
\textbf{Parameter} & \textbf{Prior Distribution} \\
\hline
$\sigma$ & $\text{Truncated Normal}(0, \sigma') \text{ with } \sigma' = 50$ \\[6pt]

$\mu$ & $\mathcal{N}(\mu_0, \tau^2)$ \\[6pt]

$\phi_a$  & $\mathcal{N}(0, \sigma''),\ \text{subject to stationarity}$ \\[6pt]

$\theta_m$ & $\mathcal{N}(0, \sigma''),\ \text{subject to invertibility}$ \\[6pt]

$D_0$  & $\mathcal{N}(\mu_0, \tau^2)$ \\[2pt]

\hline\hline
\end{tabular}
\end{table}

\subsection{Sampler Configuration}\label{sampler_config}
We use a vectorized formulation of the nested sampling algorithm developed by \citet{yallup2026nestedslicesamplingvectorized} within \texttt{blackjax}. To ensure optimal speed and compatibility with this sampler, we also built a fully vectorized ARIMA framework in Python using \texttt{JAX} \citep{jax2018github}.
\par
The sampler is initialized by specifying the model order, prior parameters, the number of live points $n_{\text{live}}$ and a random seed for the run. The process is parallelized in this sampler through the simultaneous deletion and replacement of a batch of $n_\text{delete}$ lowest-likelihood points. This parallelization parameter is set to $n_\text{delete}=50$. The \texttt{blackjax} nested sampler adopts Slice Sampling \citep{10.1214/aos/1056562461} as a default algorithm for the inner MCMC kernel of the sampler. The length of the MCMC chain run $k$ for each single replacement of the points is expressed in terms of the number of dimensions $\hat{d}$ of the parameter space. This value is conventionally set to $k=6\hat{d}$ in our work. The nested sampling run is terminated once the remaining live points are expected to contribute negligibly to the total evidence estimate. This is quantified using the ratio of the maximum possible remaining evidence contribution from the live points, $Z_{\text{live}}$, to the accumulated evidence estimate $Z$. We set a convergence criterion using the threshold value:
\begin{equation}
    \frac{Z_{\text{live}}}{Z} < 10^{-3}
\end{equation}
\subsection{ARIMA Model Selection and Fitting}\label{ARIMA_model_selection}
In this paper, we focus on evaluating the model log posterior probabilities $\log{P_i}$ on a grid of ARIMA models formed by the AR order $p$ and the MA order $q$, and a fixed $d$ order.
To ensure that the model log posterior probabilities are comparable across candidate ARIMA specifications, each model must be formulated as a complete generative description of the same observed time series. In particular, any two ARIMA models with the same $p$ and $q$ orders but different $d$ orders should be implemented without any external pre-differencing to the data. In Section \ref{results}, we demonstrate model selection across the $d$ order for a simulated ARMA process with a trend. In all other cases, considering computational costs, we fix \emph{a priori} the order of differencing $d$ using standard stationarity checks on the time series as prescribed in the Box-Jenkins methodology. In particular, statistical unit root tests such as the Augmented Dickey-Fuller (ADF) \citep{Dickey1979} and Kwiatkowski–Phillips–Schmidt–Shin (KPSS) \citep{KWIATKOWSKI1992159} Tests are used to confirm stationarity. In general, we found that the framework's preference for the $d$ order agrees with the results of the stationarity tests.
\par
A grid search using $n_{\text{live}}=100$ is first performed to identify the best model. The results are visualized on an ARMA grid by plotting a heatmap of the calculated model logarithmic posterior probabilities $\log{P_i}$ and the uncertainties estimated from the evidence calculation $\sigma_{\log{P_i}}$. This visualization is similar to \citet[fig.7]{Yu2023ApplicationRO}. From the heatmap, the model with the highest log posterior probability is picked for a second, high-resolution nested sampling run. 
\par
Further analysis is performed by dividing the time series data into training and test windows. The model with the highest posterior model probability, computed from the Bayesian evidence under uniform model priors (Equation \ref{modelposterior}), is fit to the training dataset using a high-resolution nested sampling run ($n_{\text{live}}=500$ to $1000$).
Weighted posterior samples are obtained from this run, which are then used to visualize the model fit and residuals. For this, we use \texttt{fgivenx} \citep{fgivenx} - a python package for plotting posterior line plots and predictive posteriors of functions. Using the weighted posterior samples, the posterior predictive forecasts along with their $1\sigma$, $2\sigma$ and $3\sigma$ credible regions are plotted.
\subsection{Model Validation}\label{model_validation}
The grid-search results are validated on the basis of residual analysis and forecasting accuracy. Practically, almost all ARIMA models are able to model the training window of any given time series by making one-step-ahead, in-sample predictions. On the other hand, a more accurate metric for model validation is comparing the results of out-of-sample multi-step forecasts.
\par
To confirm that the selected best-fit model has correctly captured the variability of the training dataset, we test the residuals from this fit for normality and autocorrelation. This is done by plotting residual histograms and autocorrelation plots along with performing the Ljung-Box statistical test \citep{Ljung1978} for autocorrelation. The multi-step direct forecasts for sunspot number data are also tested for accuracy by evaluating the Mean Squared Error (MSE), Root Mean Squared Error (RMSE) and Mean Absolute Error (MAE). In order to facilitate comparison with a traditional ARIMA model selection method, the Bayesian Information Criterion (BIC) heatmap is also plotted on the ARIMA grid using the ARIMA model functionality from \texttt{statsmodels}.
\par
For time series other than the sunspot number, our interest is not in forecasting performance but in the model's ability to accurately capture the variability and correlation in the given time series. Therefore, a simple residual analysis, as described previously, suffices in such cases.
\section{Results}\label{results}
This section presents the results of applying the model selection procedure to simulated and real astronomical time series data using the previously described methodologies.
\subsection{Simulated Time Series}
\begin{figure}
    \centering
    \includegraphics{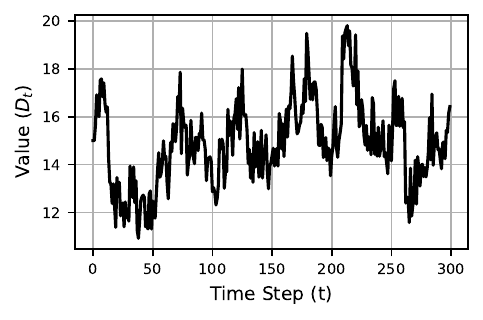}
    \caption{Artificially generated AR(2) time series of $300$ data points with $\phi_1=0.6$ and $\phi_2=0.3$. A constant intercept term of $c=1.5$ and a standard deviation of $\sigma=1.0$ associated to $\epsilon_t$ was used.}
    \label{fig:ar2_artificial}
\end{figure}

\begin{figure}
    \centering
    \includegraphics{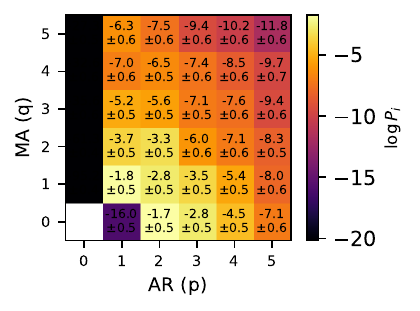}
    \caption{Heatmap of the model log posterior probabilities $P_i$ for simulated AR$(2)$ time series (Figure \ref{fig:ar2_artificial}). A hot-spot is observed at ARIMA$(2,0,0)$. The log posterior probabilities level off for higher orders as expected due to the action of Occam's penalty factor.}
    \label{fig:ar2_heatmap}
\end{figure}

\begin{figure}
    \centering
    \includegraphics{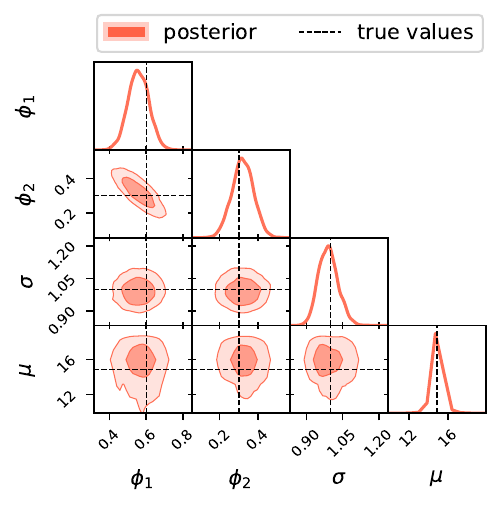}
    \caption{Posterior distributions of the AR(2) model parameters inferred from the simulated AR$(2)$ time series (Figure \ref{fig:ar2_artificial}). The 1-D kernel density estimates show well-constrained posterior densities centred near the true parameter values (indicated by the black dashed lines), thus demonstrating good recovery of the underlying process dynamics.}
    \label{fig:ar2_posteriors}
\end{figure}

The framework was first tested on a simulated AR$(2)$ time series (Figure \ref{fig:ar2_artificial}).
The grid search method correctly revealed ARIMA($2,0,0$) as the model with the highest posterior probability of $\log{P}_{\text{max}}=-1.2 \pm 0.5$ (Figure \ref{fig:ar2_heatmap}). 
To validate the methodology in this case, it is sufficient to compare the inferred posteriors to the true parameter values. A high-resolution nested sampling run of the ARIMA$(2,0,0)$ model for the time series was able to produce posterior samples close to the true values of the parameters (Figure \ref{fig:ar2_posteriors}).
\par
\begin{figure}
    \centering
    \includegraphics{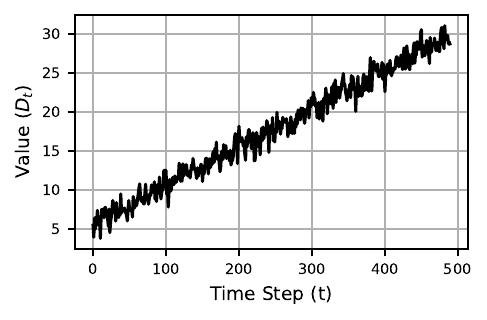}
    \caption{Artificially generated ARMA($1,1$) process of 490 data points with a linear trend. The ARMA coefficients are chosen to be $\phi_1=0.6$ and $\theta=-0.4$. The constant intercept term and standard deviation are $c=2$ and $\sigma=1$, respectively.}
    \label{sim_arima_313}
\end{figure}

To demonstrate model selection across different $d$ orders of the ARIMA model, we test our framework on a simulated ARMA($1,1$) process with a linear trend (Figure \ref{sim_arima_313}). The data were generated using the equation
\begin{equation}
    D(t) = c + \beta t+ \phi_1y_{t-1} + \theta_1 \epsilon_{t-1} + \epsilon_t
\end{equation}
with the $\beta t$ term producing the linear trend. 
Nested sampling runs were initiated on these data for ARIMA($1,d,1$) models, with $d$ ranging from $0$ to $4$. The calculated model log posterior probabilities $\log P_i$ along with their uncertainties are outlined in Table \ref{logp_d_table}.
\begin{table}
\centering
\caption{Model log-posterior probabilities $\log P_i$ and their uncertainties $\sigma_{\log P_i}$ for ARIMA($1,d,1$) fits to the data in Figure \ref{sim_arima_313}}
\begin{tabular}{c c c}
\hline\hline
$d$ & $\log P_i$ & $\sigma_{\log P_i}$ \\
\hline
0 & $-19.79499$ & $0.20408$ \\
1 & $-0.01814$  & $0.25429$ \\
2 & $-4.03641$  & $0.20032$ \\
3 & $-8.12772$  & $0.21359$ \\
4 & $-10.83405$ & $0.24954$ \\
\hline\hline
\label{logp_d_table}
\end{tabular}
\end{table}
The posterior probability is highest for ARIMA($1,1,1$) as expected, since first-order differencing is required to eliminate the linear trend in the data.

\subsection{Sunspot Number Data}
Sunspots are darker and cooler spots observed on the solar photosphere. The evolution of the number of sunspots during the sunspot cycle is found to be associated with the solar magnetic activity \citep{Hathaway2015}. It is therefore essential to analyse and forecast the sunspot number for predicting space weather and mitigating its impact on Earth.
\par
We use the yearly sunspot number data from 1700 to 2008 for our analysis. 
\begin{figure}
    \centering
    \includegraphics{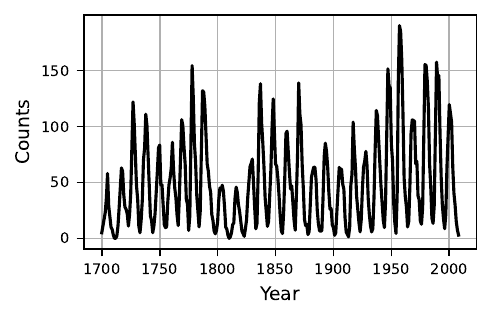}
    \caption{Yearly sunspot number data from 1700 to 2008}
    \label{fig:sunspotdata}
\end{figure}
The first 255 data points corresponding to the years from 1700 to 1954 are used as the training dataset. The results of ADF and KPSS stationarity tests, on this training data, are outlined in Table \ref{tab:stationarity_tests}. 
\begin{table}
\centering
\caption{Results of the Stationarity Tests (ADF and KPSS) on yearly sunspot number data.}
\begin{tabular}{lcc}
\hline\hline
\textbf{Statistic} & \textbf{ADF Test} & \textbf{KPSS Test} \\
\hline
Test Statistic & -2.931083 & 0.124768 \\
p-value & 0.041851 & 0.100000 \\
Lags Used & 8 & 7 \\
Number of Observations Used & 246 & 247 \\
Critical Value (1\%) & -3.457215 & 0.739000 \\
Critical Value (5\%) & -2.873362 & 0.463000 \\
Critical Value (10\%) & -2.573070 & 0.347000 \\
\hline\hline
\end{tabular}
\label{tab:stationarity_tests}
\end{table}
From the $p$-value and the critical value at 5\% significance level of both tests, it can be concluded that the time-series data is stationary. Therefore, no differencing is required and the $d$ order is set to zero for the ARIMA grid search. A grid search was carried out for ARIMA orders up to ten (Figure \ref{fig:sunspotheatmap}), which selected the ARIMA($9,0,1$) model with the posterior probability of $\log{P_{\text{max}}}=-1.3\pm0.5$. 
\begin{figure*}
    \centering
    \includegraphics{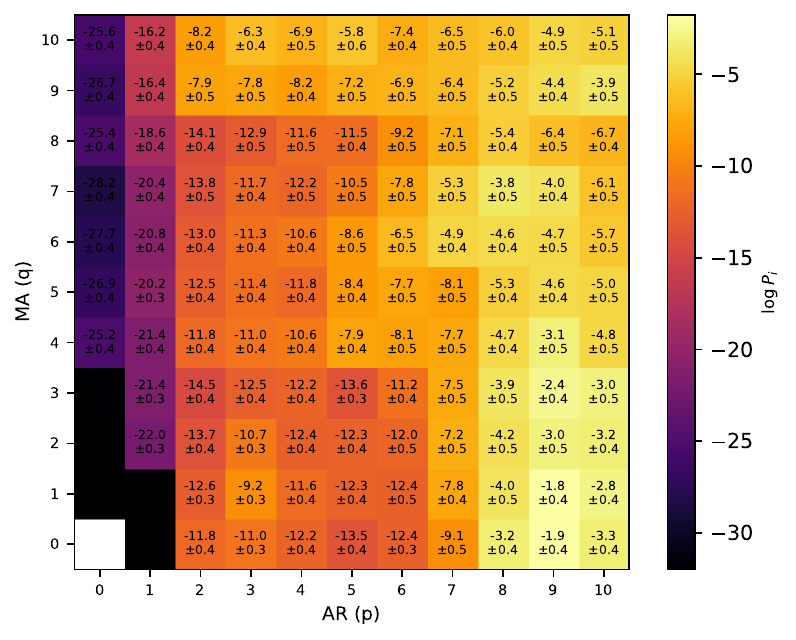}
    \caption{Heatmap of the model log posterior probabilities obtained from the nested sampling runs on yearly sunspot number data (Figure \ref{fig:sunspotdata}). The grid is annotated with model log posterior probabilities $\log{P_i}$ and their estimated uncertainties $\sigma_ {\log{P_i}}$ for reference. A clear statistical preference for higher ARIMA orders, and particularly for a higher AR order $p$ can be seen, with ARIMA$(9,0,1)$ having the highest model posterior probability.}
    \label{fig:sunspotheatmap}
\end{figure*}

\begin{figure}
    \centering
    \includegraphics{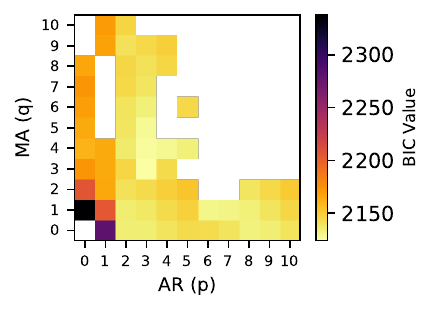}
    \caption{Heatmap of the Bayesian Information Criterion (BIC) values for ARIMA model fit on yearly sunspot number data. Missing data indicate models for which the maximum likelihood estimation failed to converge. The lowest value of BIC is observed for ARIMA($3,0,3$)}
    \label{fig:sunspots_bic}
\end{figure}

The maximum likelihood estimation method adopted by the ARIMA model function in \texttt{statsmodels} failed to converge for multiple higher-order ARIMA models (see Figure \ref{fig:sunspots_bic}). ARIMA$(3,0,3)$ was picked as the best-fit model using this method by minimizing the Bayesian Information Criterion (BIC). To confirm this is not an artefact of poor initialisation, we additionally ran \texttt{auto\_arima} from the \texttt{pmdarima} library with a full grid search up to $p=q=10$ and BIC as the selection criterion. This returned ARIMA$(3,0,2)$ as the best-fit model and also reported convergence failures at some higher-order models, consistent with the \texttt{statsmodels} result. The close model preference between two independent MLE implementations suggests that this disagreement with our nested sampling result is a fundamental limitation of likelihood optimisation in high-dimensional ARMA parameter spaces. As demonstrated in Appendix A, ARIMA$(9,0,1)$ selected by nested sampling achieves substantially better performance metrics on the held-out test window than ARIMA$(3,0,3)$, providing empirical validation that the Bayesian evidence identifies a richer model structure in this case than MLE-based methods.
\par
The preference for a high AR order of $p=9$ in the heatmap data (Figure \ref{fig:sunspotheatmap}) can be physically motivated. Solar magnetic activity is known to operate across a hierarchy of timescales. The dominant periodicity is the Schwabe cycle of approximately 11 years, but significant structure also exists on shorter timescales associated with active region emergence, differential rotation, and the stochastic nature of flux tube emergence at the solar surface. In an AR model, the order $p$ determines the memory depth of the process in units of the sampling interval. With annual sampling, an AR order of 9 therefore captures dependencies extending approximately 9 years into the past, commensurate with the rise phase of the solar cycle. The single MA term ($q = 1$) is consistent with a short-memory noise process superimposed on this long-range autoregressive structure. 

\begin{figure*}
    \centering
    \includegraphics{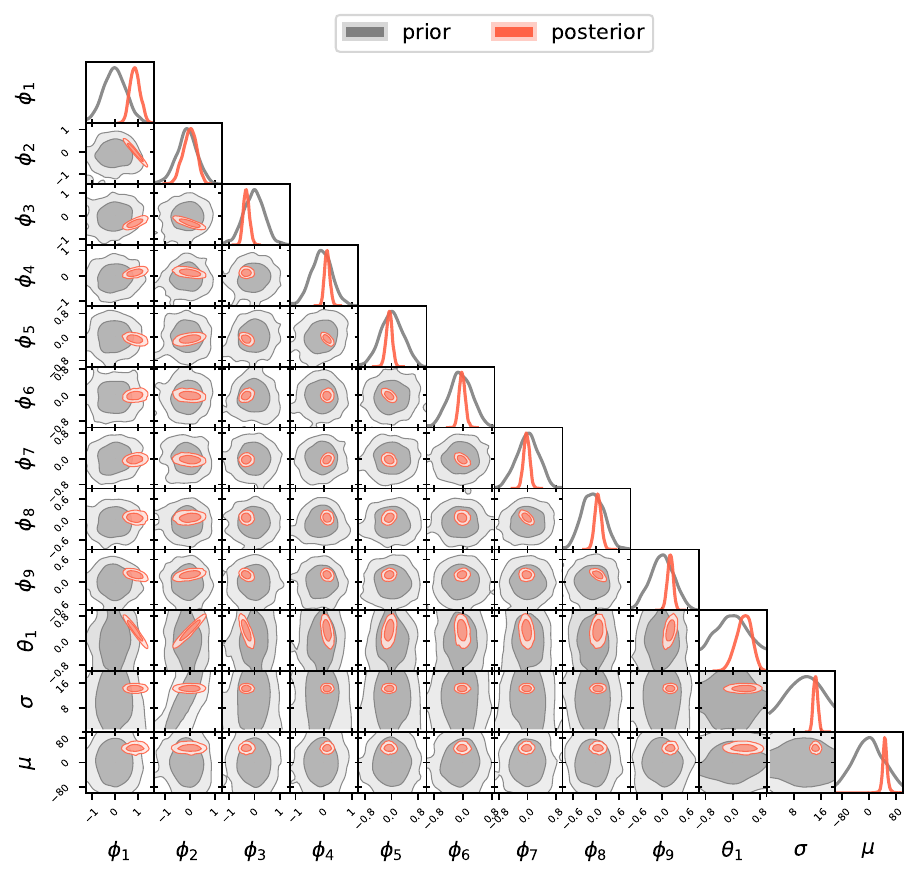}
    \caption{Posterior corner plot of ARIMA($9,0,1$) model parameters obtained from the nested sampling run on the yearly sunspot number data (Figure \ref{fig:sunspotdata}). Orange curves represent the marginalized posterior distributions, while the gray curves show the corresponding prior distributions (Table \ref{tab:prior_table}) adopted for each parameter. The sharp posterior peaks indicate well-constrained AR and MA coefficients, with most of the posterior mass concentrated within the stationary and invertible regions of parameter space. The information gain between prior and posterior, measured by the Kullback–Leibler divergence, is $D_{KL}=19$ nats, demonstrating a substantial concentration of probability mass due to the data.}
    \label{fig:full_posteriors}
\end{figure*}

\begin{figure}
    \centering
    \includegraphics{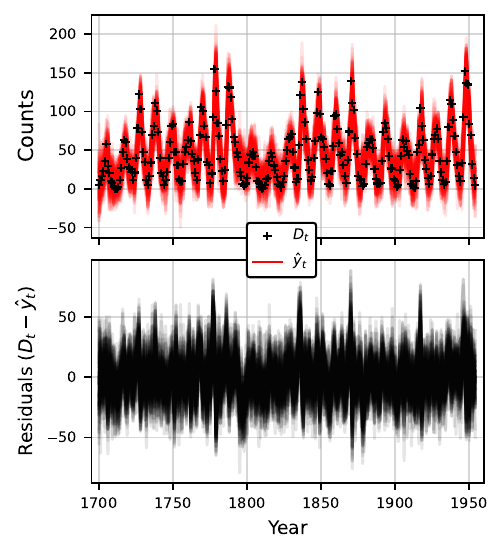}
    \caption{Results of ARIMA$(9,0,1)$ model fit to the training window, corresponding to the years of from 1700 to 1954, of the yearly sunspot number data (Figure \ref{fig:sunspotdata}). The upper panel shows the observed sunspot counts (in black) and the model fit curves (in red) obtained using $500$ weighted posterior samples of the ARIMA parameters, while the lower panel displays the corresponding residuals. The residuals fluctuate around zero with no strong temporal structure, indicating that the fitted model captures most of the systematic variability in the data. }
    \label{fig:sunspot_fit_residuals}
\end{figure}

\begin{figure}
    \centering
    \includegraphics{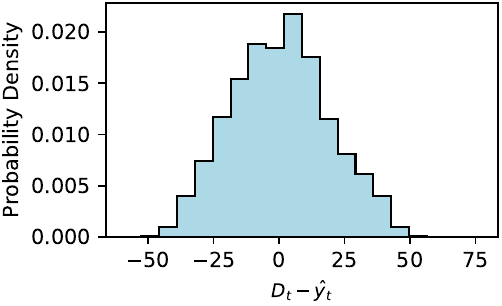}
    \caption{A histogram of the pooled residuals from the ARIMA$(9,0,1)$ fit in Figure \ref{fig:sunspot_fit_residuals} to yearly sunspot number data.}
    \label{fig:sunspot_resid_hist}
\end{figure}

We choose ARIMA$(9,0,1)$ for fitting the data using a high-resolution ($n_\text{live}=1000$) nested sampling run. The marginalized posterior distributions of ARIMA parameters obtained from this fit are shown in Figure \ref{fig:full_posteriors}.
The model fit and residuals are shown in Figure \ref{fig:sunspot_fit_residuals}. 
We note that the model occasionally produces negative predicted sunspot counts, which are unphysical. This is a known limitation of applying a Gaussian likelihood to strictly non-negative count data. A more robust treatment would adopt a truncated Gaussian or a negative binomial likelihood. It can be seen that the residual sequence symmetrically fluctuates around zero. 
A histogram of the pooled residuals from different posterior samples is shown in Figure \ref{fig:sunspot_resid_hist}.
We normalized the pooled residuals by the posterior mean of the inferred standard deviation $\sigma$. These normalized forecast residuals have a mean of $0.01$ and a standard deviation of $1.29$, indicating that the predictive uncertainty scale inferred by the model is consistent with the observed forecast errors.

\begin{figure}
    \centering
    \includegraphics{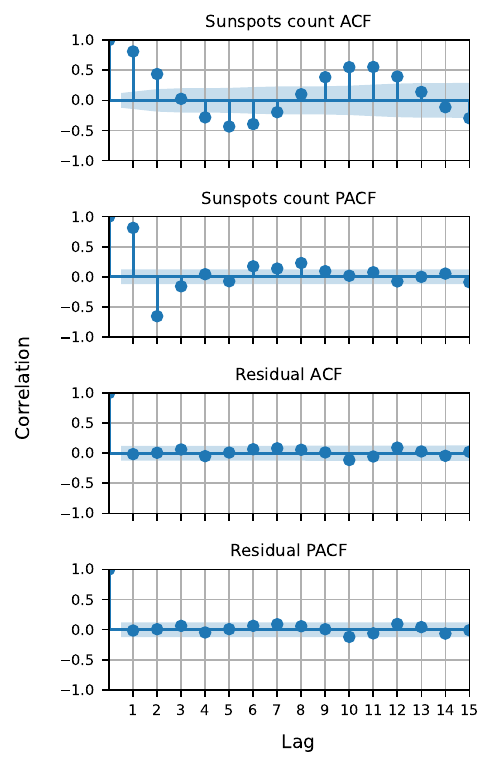}
    \caption{Autocorrelation (top) and Partial Autocorrelation (bottom) function plots of the mean residuals from the ARIMA$(9,0,1)$ fit to the yearly sunspot number data (Figure \ref{fig:sunspot_fit_residuals}). Both functions lie within the 95\% confidence bounds across all lags, indicating that the residuals are effectively uncorrelated.}
    \label{fig:acf_sunspots}
\end{figure}

We also used the time-ordered residuals of the fit, obtained from the posterior means of the parameters, to further check for any autocorrelation in the residuals. The Autocorrelation function (ACF) and Partial Autocorrelation function (PACF) plots (Figure \ref{fig:acf_sunspots}) suggest no significant autocorrelation in the residual time series. 
This is further confirmed from the results of the Ljung-Box test, up to lag $10$, on the time-ordered residuals (Table \ref{tab:ljung_box_test}). 
\begin{table}
\centering
\caption{Ljung--Box(LB) test p-values for residual autocorrelation of sunspot number data.}
\label{tab:ljung_box_test}
\begin{tabular}{ccc}
\hline\hline
\textbf{Lag} & \textbf{p-value} \\
\hline
1   & 0.841791 \\
2   & 0.978972 \\
3    & 0.728602 \\
4    & 0.819598 \\
5    & 0.905064 \\
6    & 0.785299 \\
7    & 0.624773 \\
8   & 0.655938 \\
9   & 0.746816 \\
10  & 0.535520 \\
\hline\hline
\end{tabular}
\end{table}

The results show that all $p$-values exceed $0.05$ across lags $1$-$10$. Hence, we fail to reject the null hypothesis of no serial correlation in the residuals at the $5\%$ significance level. The residuals are therefore consistent with a white noise sequence, indicating that the ARIMA$(9,0,1)$ fit has effectively captured the temporal structure of our training data.
\par
\begin{figure}
    \centering
    \includegraphics{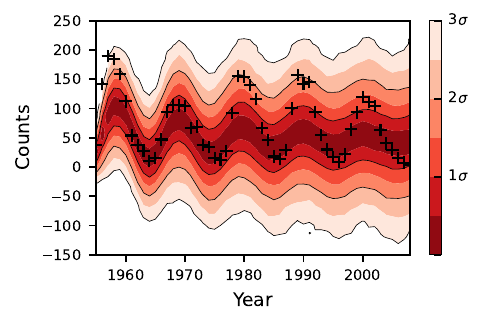}
    \caption{Posterior predictive forecasts of the yearly sunspot number for the years from 1954 to 2008. The forecasts are obtained using $5000$ weighted posterior samples from the ARIMA$(9,0,1)$ fit (Figure \ref{fig:sunspot_fit_residuals}). Shaded contours denote the $1\sigma$, $2\sigma$ and $3\sigma$ credible regions of the predictive posterior distribution $P(\hat{y}_t|t,D_t)$.}
    \label{fig:arima9_forecast}
\end{figure}

Finally, we show in Figure \ref{fig:arima9_forecast}, the results of direct multi-step forecasting of the sunspot number data from 1954 to 2008, using an ARIMA$(9,0,1)$ fit.
In Appendix \ref{sunspots_comparison}, the forecasting performance of ARIMA$(9,0,1)$ is compared to ARIMA$(3,0,3)$. Although the latter model is favoured by the Bayesian Information Criterion (BIC) (Figure \ref{fig:sunspots_bic}), it yields a significantly lower model log posterior probability compared to ARIMA$(9,0,1)$. 
\subsection{Kepler and TESS light curves}
The Kepler and TESS (Transiting Exoplanet Survey Satellite) photometric light curves act as ideal datasets for testing our ARIMA-Nested Sampling framework owing to their long and regular cadences with minimal systematics noise. In this subsection, we present the results of applying our model selection methodology to photometric light curves from the Kepler and TESS missions. We focus on analysing the light curves of three distinct astrophysical systems: a low-luminosity red giant -- KIC 12008916, an active G-type star hosting a hot Jupiter -- Kepler 17, and two distinct quasars observed by TESS -- 3C 273 and S4 0954+65.
\subsubsection*{KIC 12008916}
KIC 12008916 is a low-luminosity red giant in the original Kepler field and is one of the benchmark stars in the Kepler asteroseismic sample \citep{https://doi.org/10.1002/asna.201612371}. The star has a well-resolved, solar-like oscillation spectrum. Its light curve exhibits stochastic variability which is stable and stationary over longer timescales, providing an ideal test-bed for ARIMA modelling. 
\par
\begin{figure}
    \centering
    \includegraphics[width=\linewidth]{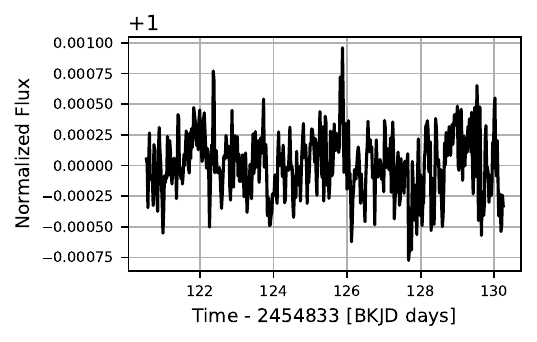}
    \caption{Kepler long-cadence light curve of KIC 12008916, normalized and shown over a 10-day segment. The star exhibits the characteristic stochastic, solar-like oscillations of a low-luminosity red giant.}
    \label{kic_lc}
\end{figure}
\begin{figure*}
    \centering
    \includegraphics{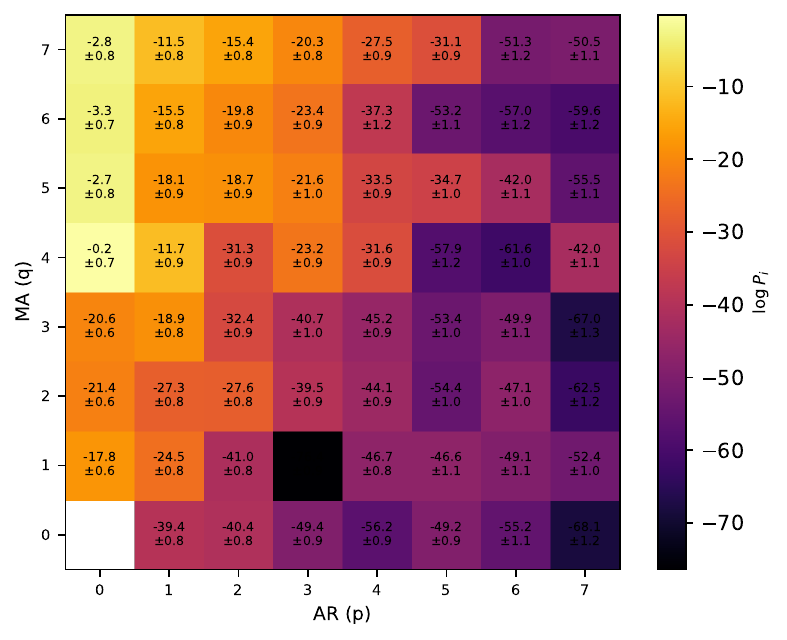}
    \caption{Heatmap of the ARIMA models' log posterior probabilities obtained from the nested sampling runs on KIC 12008916 data (Figure \ref{kic_lc}). A statistical preference for purely moving average models is seen, with the highest log posterior probability for ARIMA$(0,0,4)$.}
    \label{kic_heatmap}
\end{figure*}

We use light curve data from the Kepler Quarter $00$, corresponding to the year of 2009 (Figure \ref{kic_lc}).
The Pre-search Data Conditioning Simple Aperture Photometry (PDCSAP) flux was used for this analysis, which was found to be significantly autocorrelated, as seen from the Autocorrelation Function (ACF) and Partial Autocorrelation Function (PACF) plots in the top two panels of Figure \ref{kic_acf}. The results of the ADF and KPSS stationarity tests confirmed the time series data to be stationary. 
\begin{table}
\centering
\caption{Results of Augmented Dickey-Fuller (ADF) and KPSS Stationarity Tests for KIC 12008916 light curve.}
\begin{tabular}{lcc}
\hline\hline
\textbf{Statistic} & \textbf{ADF Test} & \textbf{KPSS Test} \\
\hline
Test Statistic & -4.620294 & 0.100269 \\
p-value & 0.000118 & 0.100000 \\
Lags Used & 9 & 10 \\
Number of Observations Used & 459 & -- \\
Critical Value (1\%) & -3.444677 & 0.739000 \\
Critical Value (5\%) & -2.867857 & 0.463000 \\
Critical Value (10\%) & -2.570135 & 0.347000 \\
\hline\hline
\end{tabular}
\end{table}
An ARIMA grid search, with $d=0$ fixed, revealed ARIMA$(0,0,4)$ as the best-fit model ($\log{P_{\text{max}}}=-0.18 \pm0.67$). The results of the grid search are shown in Figure \ref{kic_heatmap}.

\begin{figure}
    \centering
    \includegraphics[width=\linewidth]{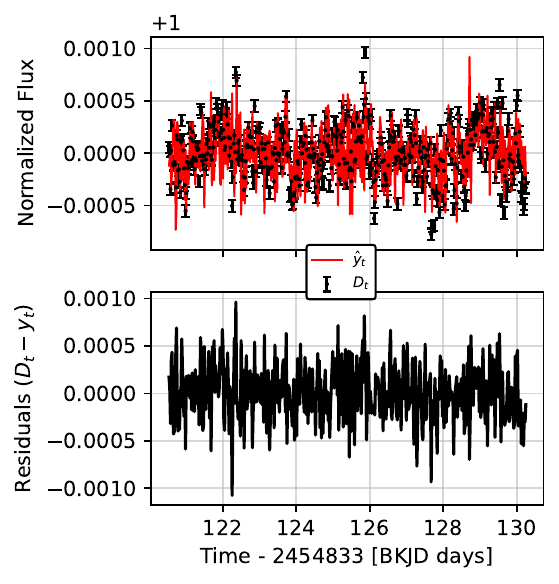}
    \caption{Results of ARIMA$(0,0,4)$ model fit to the KIC 12008916 light curve in Figure \ref{kic_lc}. Top panel shows the light curve data (in black) and the model fit curve (in red) obtained from the mean of the posterior samples. Bottom panel shows the residuals obtained after subtracting the mean model fit from the data. The residual time series show no temporal structure and is characteristic of a white noise sequence.}
    \label{fig:kic_12008916_fit}
\end{figure}

\begin{figure}
    \centering
    \includegraphics[width=\linewidth]{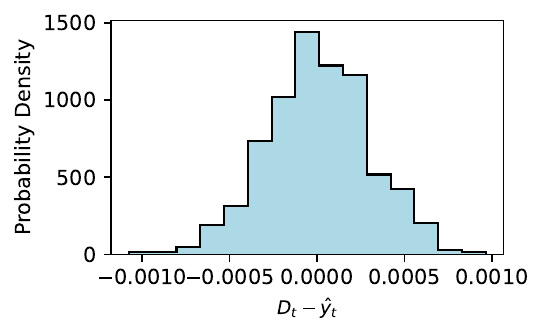}
    \caption{A histogram of the residuals in Figure \ref{fig:kic_12008916_fit}, obtained from the ARIMA$(0,0,4)$ model fit to the light curve data of KIC 12008916.}
    \label{fig:kic12008916_resid_hist}
\end{figure}

\begin{figure}
    \centering
    \includegraphics{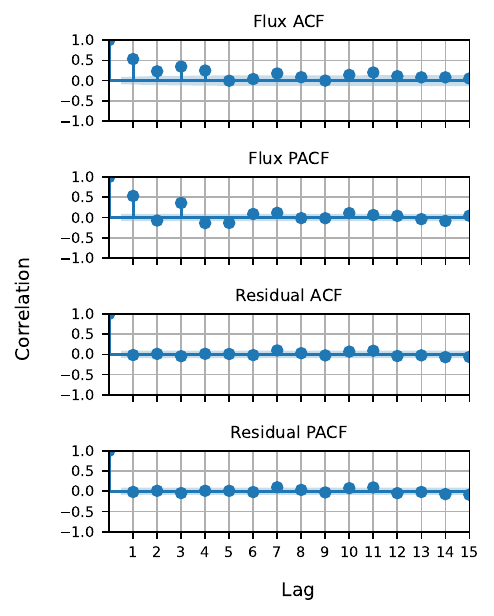}
    \caption{Autocorrelation (ACF) and Partial autocorrelation (PACF) plots for the light curve of KIC 12008916 (top) and the residuals (bottom) obtained after subtracting the best fit ARIMA$(0,0,4)$ model from the light curve data. The residual ACF/PACF plots show that the model fit has reduced most of the correlation present in the light curve data.}
    \label{kic_acf}
\end{figure}

The ARIMA$(0,0,4)$ model was fit to the data using a high-resolution ($n_{\text{live}}=500$) nested sampling run and the means of the posterior samples were used to obtain time-ordered residuals (see Figure \ref{fig:kic_12008916_fit}).
A histogram of these residuals is shown in Figure \ref{fig:kic12008916_resid_hist}.
The autocorrelation and partial autocorrelation plots of these residuals are also plotted in Figure \ref{kic_acf}. We observe no significant autocorrelation in the time-ordered residuals, as seen from these plots. 
The results of the Ljung-Box test, up to lag $10$, on the time-ordered residuals (see Table \ref{tab:ljung_box_test_kic}) also show no correlation at the 5\% significance level. The ARIMA$(0,0,4)$ model has therefore successfully modelled the stochastic variability in the original light curve of KIC 12008916.
\begin{table}
\centering
\caption{Ljung--Box (LB) test p-values for residual autocorrelation of KIC 12008916 data.}
\label{tab:ljung_box_test_kic}
\begin{tabular}{cc}
\hline\hline
\textbf{Lag} & \textbf{p-value} \\
\hline
1   & 0.720115 \\
2   & 0.906690 \\
3   & 0.765711 \\
4   & 0.871563 \\
5   & 0.936565 \\
6   & 0.963805 \\
7   & 0.539844 \\
8   & 0.602335 \\
9   & 0.667972 \\
10  & 0.540848 \\
\hline\hline
\end{tabular}
\end{table}

\subsubsection*{Kepler 17}
The detection of transiting exoplanets is sometimes impeded due to the presence of autocorrelated noise in the light curves \citep{10.1111/j.1365-2966.2006.11012.x}. This autocorrelated variability can arise from activity of the host star or also from systematics. Autoregressive modelling of the correlated noise is effective in such cases for enhancing the signal-to-noise ratio of transit detections \citep{Melton_2024}.
\par
\begin{figure}
    \centering
    \includegraphics{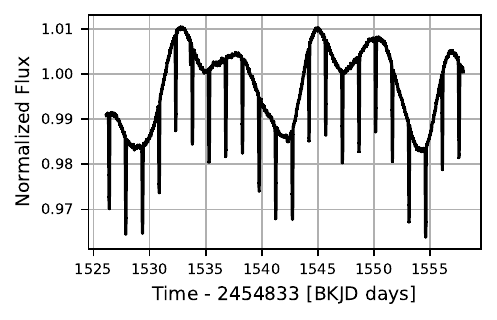}
    \caption{The normalized photometric light curve of Kepler-17, from Kepler Quarter 16, showing periodic transit dips caused by the planet Kepler-17b. The transit dips are superimposed on a quasi-periodic flux resulting primarily from spot modulation -- a combined effect of starspots and stellar rotation.}
    \label{fig:kepler17_lc}
\end{figure}
\begin{figure*}
    \centering
    \includegraphics{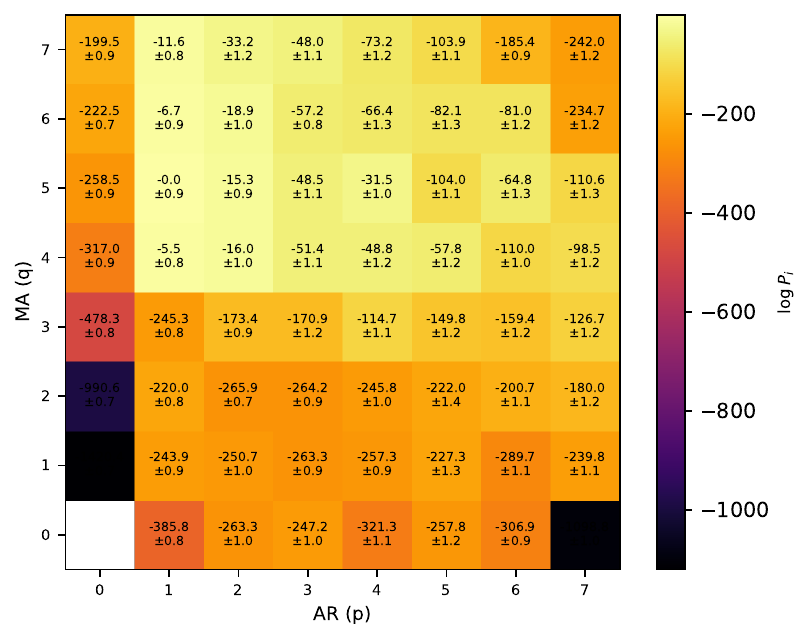}
    \caption{Heatmap of the ARIMA model's log posterior probabilities obtained from the nested sampling runs on the Kepler 17 data (Figure \ref{fig:kepler17_lc}). The differencing order is fixed to $d=1$. The grid search shows ARIMA$(1,1,5)$ as the model with highest log posterior probability.}
    \label{fig:k17_heatmap}
\end{figure*}

Here, we demonstrate a proof-of-concept by applying the ARIMA-Nested Sampling framework to the photometric light curve data of Kepler 17 -- a very active, main-sequence star of the spectral class G. The star was confirmed to host a transiting hot Jupiter -- Kepler 17b \citep{Désert_2011}, with a period of approximately $1.5$ days. For this analysis, we used the normalized PDCSAP flux of the Kepler light curve, corresponding to the Kepler Quarter 16 and year 2013 (Figure \ref{fig:kepler17_lc}). The original light curve had a sampling gap of around $28$ minutes in the data while the cadence was around $0.86$ minutes. Figure \ref{fig:kepler17_lc} was obtained by binning the data into time-bins of $0.0211$ days ($30.4$ minutes), to render the sampling rate uniform. By doing so, we focus on the stellar variability exhibited on longer scales but lose information about the short-timescale stochastic variability.
This light curve exhibits a significant transit-depth from the exoplanet and spot modulation due to the stellar activity which also reflects the rotation rate of the star \citep{k17_rotation}. The data was tested for stationarity using the ADF and KPSS tests, and was found to be non-stationary according to the ADF test (see Table \ref{tab:k17_stationarity}). The time series was rendered stationary after first-order differencing $d=1$.
\begin{table}
\centering
\caption{Results of Augmented Dickey-Fuller (ADF) and KPSS Stationarity Tests for Kepler 17 light curve.}
\label{tab:k17_stationarity}
\begin{tabular}{lcc}
\hline\hline
\textbf{Statistic} & \textbf{ADF Test} & \textbf{KPSS Test} \\
\hline
Test Statistic & -2.438782 & 0.431541 \\
p-value & 0.131088 & 0.063560 \\
Lags Used & 14 & 24 \\
Number of Observations Used & 1493 & -- \\
Critical Value (1\%) & -3.434738 & 0.739000 \\
Critical Value (5\%) & -2.863478 & 0.463000 \\
Critical Value (10\%) & -2.567802 & 0.347000 \\
\hline\hline
\end{tabular}
\end{table}
The grid search was performed on this differenced data up to a maximum AR and MA order of $7$. The results of this grid search, shown in Figure \ref{fig:k17_heatmap}, reveal ARIMA$(1,1,5)$ as the model with the highest posterior probability ($\log{P_{\text{max}}}=-0.0054 \pm0.88$). 

A high-resolution nested sampling run ($n_{\text{live}}=500$) was performed to fit the ARIMA$(1,1,5)$ model to the data, and as before, time-ordered residuals using posterior means were obtained.
\begin{figure}
    \centering
    \includegraphics[width=\linewidth]{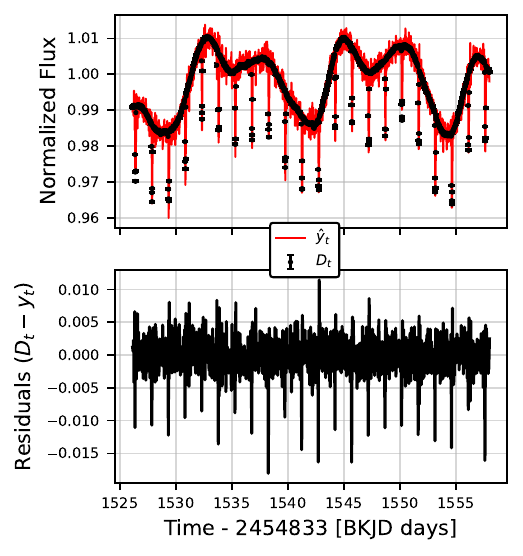}
    \caption{Results of the ARIMA$(1,1,5)$ model fit to the normalized light curve of Kepler 17 (Figure \ref{fig:kepler17_lc}). Top panel shows the light curve data (in black) and the model fit curve (in red) obtained from the mean of the posterior samples. Bottom panel shows the corresponding residual time series. The residuals depict clear periodic spikes corresponding to the transit dips of Kepler 17-b.}
    \label{fig:k17_fit}
\end{figure}
The residual time series still depicts a periodic structure resulting from the planet transits. This shows that even though we have not masked the transits during the fit, the ARIMA model comparison and fit is not completely biased by the transits. This preservation of the unmasked transient features within the residuals aligns closely with the principles of Bayesian anomaly detection established by \citet{leeney2025bayesiananomalydetectionia}. 

\begin{figure}
    \centering
    \includegraphics{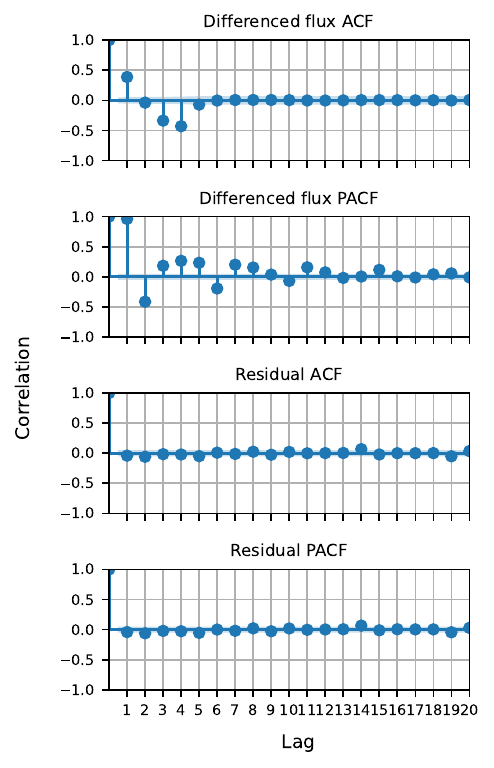}
    \caption{Autocorrelation and Partial autocorrelation plots for the light curve of Kepler 17 (top) and the residuals (bottom) obtained after subtracting the best fit ARIMA$(1,1,5)$ model from the light curve data.}
    \label{fig:k17_acf}
\end{figure}

For brevity, we will not show the histogram of these residuals, but inspect the ACF/PACF plots as shown in Figure \ref{fig:k17_acf}.
These plots show that the model has removed most of the correlation present in the original data. Table \ref{tab:ljung_box_test_k17} shows statistically significant autocorrelation at lags 2, 3, and 5, indicating that the ARIMA$(1,1,5)$ model has not fully whitened the residuals in the strict sense required for a complete ARIMA fit. This residual correlation is, however, physically expected. The periodic transit signature of Kepler-17b at a period of approximately 1.5 days introduces a deterministic, non-stochastic signal component that ARIMA, as a purely stochastic model, is not designed to efficiently absorb. A more complete, future treatment would involve masking the transit epochs prior to ARIMA fitting, or employing a hybrid model that explicitly accounts for both the stochastic stellar component and the periodic planetary signal. 

\begin{figure}
    \centering
    \includegraphics[width=\linewidth]{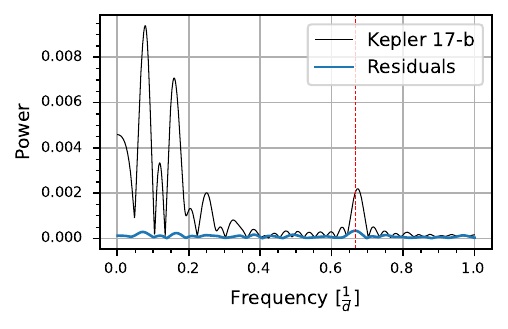}
    \caption{Lomb-Scargle periodograms of the light curve data (black) and ARIMA$(1,1,5)$ fit residuals (blue) from Figure \ref{fig:k17_fit}. The periodogram of the original light curve shows multiple peaks associated with stellar variability, with the strongest peak corresponding to the stellar rotation period of $\sim11$ days. The peak near $0.66/\mathrm{day}$ corresponds to the transit period of Kepler-17b ($\sim1.5$ days). In the residual periodogram, the stellar variability is significantly suppressed while the planetary transit signal remains relatively prominent.}
    \label{fig:k17_pg}
\end{figure}

\begin{table}
\centering
\caption{Ljung--Box (LB) test p-values for residual autocorrelation of Kepler 17 light curve.}
\label{tab:ljung_box_test_k17}
\begin{tabular}{cc}
\hline\hline
\textbf{Lag} & \textbf{p-value} \\
\hline
1   & 0.116686 \\
2   & 0.021184 \\
3   & 0.044350 \\
4   & 0.067597 \\
5   & 0.029983 \\
6   & 0.051579 \\
7   & 0.079548 \\
8   & 0.091245 \\
9   & 0.103069 \\
10  & 0.121856 \\
\hline\hline
\end{tabular}
\end{table}
To further determine whether this fit has removed the periodic signal due to stellar activity, we inspect the periodograms of the light curve and the residuals in Figure \ref{fig:k17_pg}. Figure \ref{fig:k17_pg} shows the Lomb-Scargle periodograms of the light curve data and the residual from the ARIMA model fit, evaluated on a frequency grid of $0.001/\text{day}$ to $1/\text{day}$. 
\par
The light curve periodogram has many peaks at different frequencies. The period at maximum power corresponds to the known stellar rotation period of approximately $11$ days. The peak at frequency corresponding to $1.5$ days, marked by the red dashed line, is from the periodic transit signals of Kepler 17b. We estimate the relative prominence of the transit signal by computing the ratio between the periodogram power at the transit frequency and the power at the stellar rotation frequency. This ratio from the light curve periodogram was calculated to be around $0.21$. For the residual's periodogram, the period at maximum power corresponded to the transit period of $1.5$ days. A similar calculation of the relative strength of the transit signal with respect to the power at the stellar rotation peak yielded a value of $1.27$. This ratio increase from $0.21$ to $1.27$ indicates that the ARIMA$(1,1,5)$ fit has modelled the stellar rotation signal while relatively preserving the transit periodicity.
\subsubsection*{3C 273 and S4 0954+65}
Quasars represent the brightest and most extreme objects in the universe and fall under the broader classification of Active Galactic Nuclei (AGN). The photometric variability of AGNs in multiple wavelengths is an active area of research. The study of their variability offers insights into the physical processes governing accretion and the activity of the AGN's central engine. ARIMA can be used to effectively model the stochastic variability depicted by the light curves of such quasars \citep{10.1093/mnras/stv004,Belete_2019}.
\par
Here, we present the results of applying our ARIMA-Nested Sampling framework to two quasars -- 3C 273 and S4 0954+65. The former was identified in the Third Cambridge Catalogue of Radio Sources \citep{Archer_Baldwin_Edge_Elsmore_Scheuer_Shakeshaft_1959}. It is the brightest quasar in the optical apparent magnitude and the first to be spectroscopically identified. S4 0954+65 is a blazar of the BL Lac subtype, discovered in the S4 radio survey \citep{1977Natur.268..405C}. Both quasars were extensively observed by the TESS mission over multi-day observing sectors. In this analysis, we use data from the Sector 46 (year 2021) for 3C 273 and Sector 20 (year 2019) for S4 0954+65. The data was obtained and processed using the MIT Quick Look Pipeline (QLP) \citep{Huang_2020A,Huang_2020B} for 3C 273 and using the TESS Science Processing Operations Center (TESS-SPOC) pipeline \citep{Jenkins2016} for S4 0954+65.
\par
\begin{figure*}
    \centering
    \includegraphics{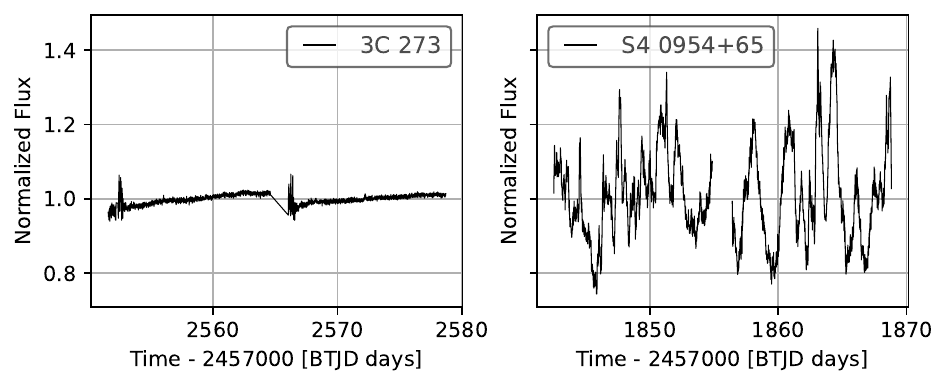}
    \caption{TESS normalized light curves of the quasar 3C 273 (left) and the blazar S4 0954+65 (right) plotted on a shared y-axis. S4 0954+65 light curve is obtained by binning the original light curve in time with a bin size of $20.16$ minutes. From the plots, S4 0954+65 exhibits significantly more optical variability than 3C 273.}
    \label{fig:quasars_lc}
\end{figure*}

Figure \ref{fig:quasars_lc} shows the normalized light curves of 3C 273 and S4 0954+65. The latter is produced by binning the original flux data into time bins of $20.16$ minutes while the cadence of the original light curve data was around $1.87$ minutes.
There is a single significant sampling gap in both light curves. Therefore, for the grid search and rest of the analysis, we use the data before the sampling gaps as our training dataset. The light curve of 3C 273 shows little to no variability compared to S4 0954+65, except the two jitters due to pointing instabilities. These bad values are also clipped from our training data. The results of the ADF and KPSS stationarity tests (see Table \ref{tab:stationarity_quasar} and Table \ref{tab:stationarity_blazar}) indicated both time series to be non-stationary. This non-stationarity is also apparent from their light curves, where the 3C 273 light curve exhibits a linear trend and S4 0954+65 light curve shows heteroscedasticity. 
\begin{table}
\centering
\caption{Results of Augmented Dickey-Fuller (ADF) and KPSS Stationarity Tests for 3C 273 light curve.}
\label{tab:stationarity_quasar}
\begin{tabular}{lcc}
\hline\hline
\textbf{Statistic} & \textbf{ADF Test} & \textbf{KPSS Test} \\
\hline
Test Statistic & -1.623081 & 5.853438 \\
p-value & 0.471114 & 0.010000 \\
Lags Used & 23 & 26 \\
Number of Observations Used & 1586 & -- \\
Critical Value (1\%) & -3.434480 & 0.739000 \\
Critical Value (5\%) & -2.863364 & 0.463000 \\
Critical Value (10\%) & -2.567741 & 0.347000 \\
\hline\hline
\end{tabular}
\end{table}

\begin{figure}
    \centering
    \includegraphics{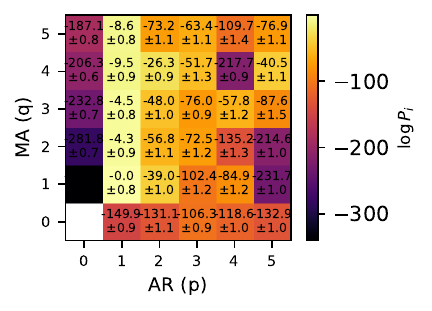}
    \caption{Heatmap of the ARIMA model's log posterior probabilities obtained from the nested sampling runs on the 3C 273 light curve data. The differencing order is fixed to $d=1$. The grid search shows ARIMA$(1,1,1)$ as the best-fit model with highest log posterior probability.}
    \label{fig:quasar_heatmap}
\end{figure}

\begin{figure*}
    \centering
    \includegraphics{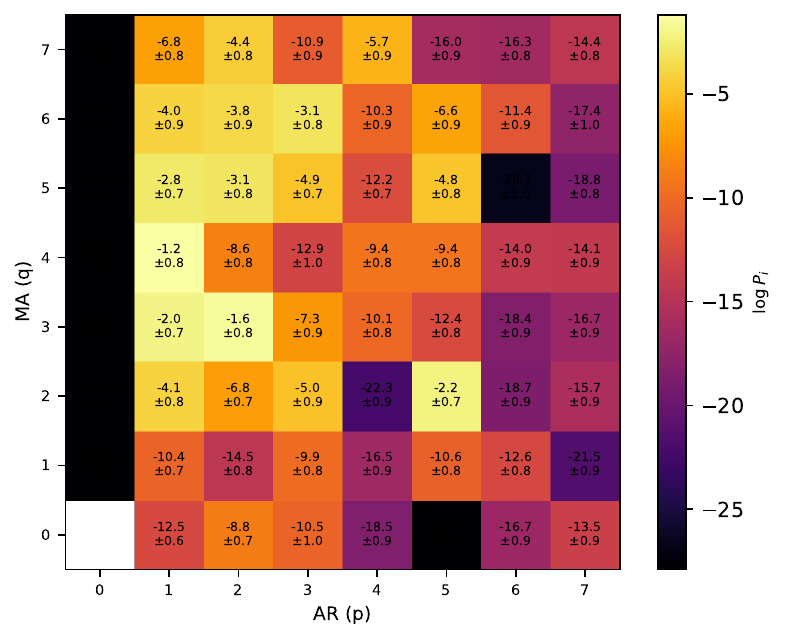}
    \caption{Heatmap of the ARIMA model's log posterior probabilities obtained from the nested sampling runs on the S4 0954+65 light curve data. The differencing order is fixed to $d=1$. The grid search shows ARIMA$(1,1,4)$ as the best-fit model with highest log posterior probability.}
    \label{fig:blazar_heatmap}
\end{figure*}

\begin{figure}
    \centering
    \includegraphics{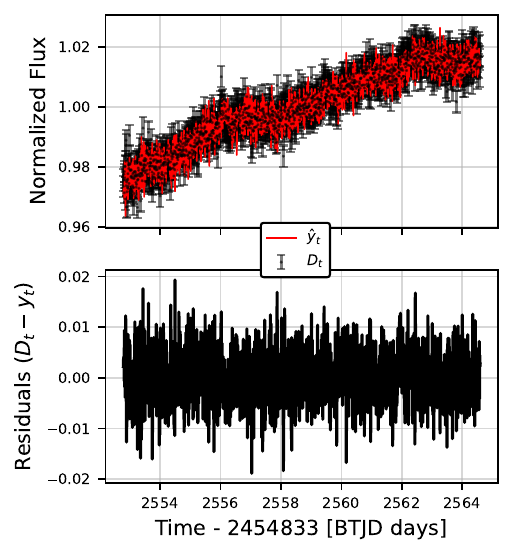}
    \caption{Results of the ARIMA$(1,1,1)$ model fit to the normalized light curve of 3C 273 (Figure \ref{fig:quasars_lc}).}
    \label{fig:quasar_fit}
\end{figure}

\begin{figure}
    \centering
    \includegraphics{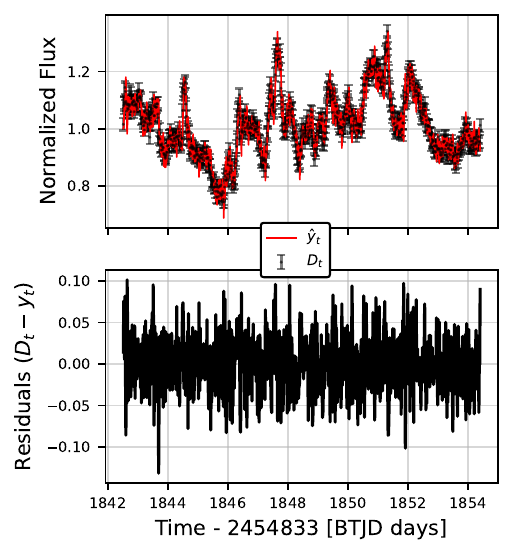}
    \caption{Results of the ARIMA$(1,1,4)$ model fit to the normalized light curve of S4 0954+65 (Figure \ref{fig:quasars_lc}).}
    \label{fig:blazar_fit}
\end{figure}

\begin{figure*}
    \centering
    \includegraphics{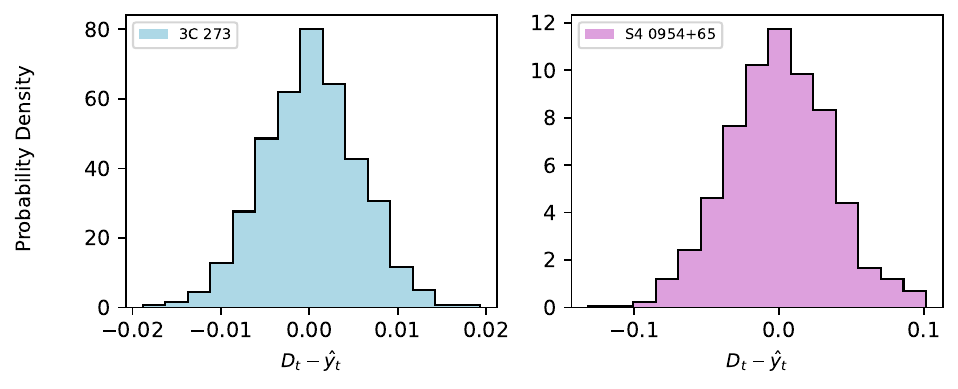}
    \caption{Histograms of the residuals of 3C 273 (left) and S4 0954+65 (right) obtained after subtracting the best fit ARIMA models.}
    \label{fig:quasar_resid_hist}
\end{figure*}

\begin{figure}
    \centering
    \includegraphics{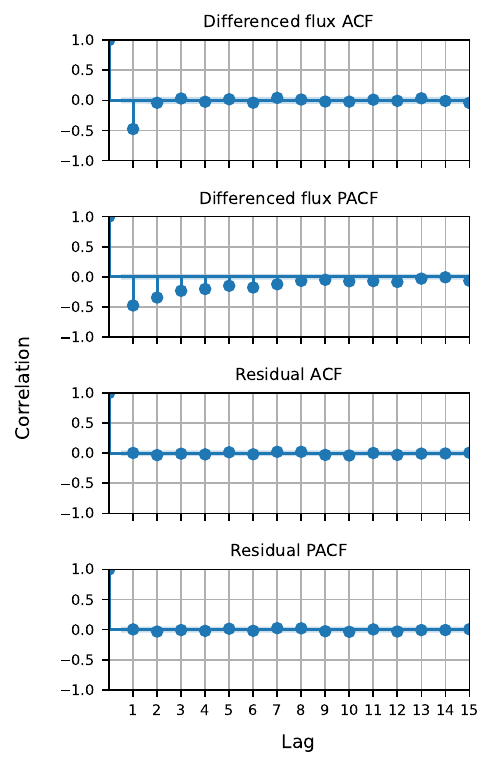}
    \caption{Autocorrelation (ACF) and Partial autocorrelation (PACF) plots for the light curve of 3C 273 (top) and the residuals (bottom) obtained after subtracting the best fit ARIMA$(1,1,1)$ model from the light curve data. The residual ACF/PACF plots show that the model fit has reduced most of the correlation present in the light curve data.}
    \label{fig:quasar_acf}
\end{figure}

\begin{figure}
    \centering
    \includegraphics{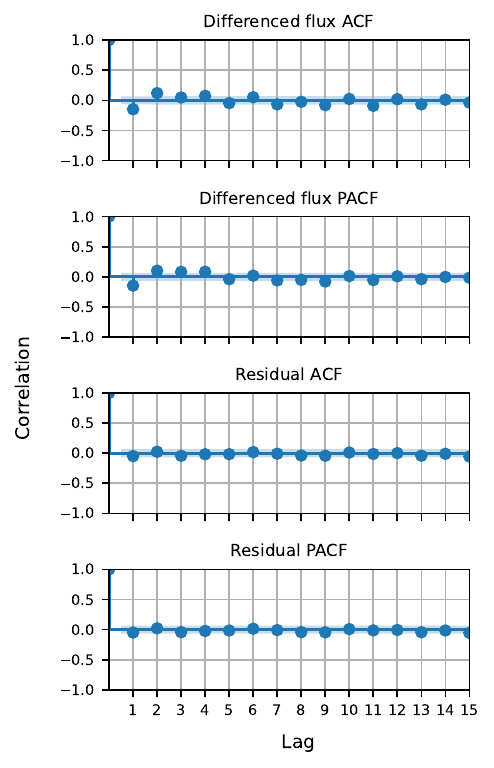}
    \caption{Autocorrelation (ACF) and Partial autocorrelation (PACF) plots for the differenced light curve of S4 0954+65 (top) and the residuals (bottom) obtained after subtracting the best fit ARIMA$(1,1,4)$ model from the light curve data. The residual ACF/PACF plots show that the model fit has reduced most of the correlation present in the light curve data.}
    \label{fig:blazar_acf}
\end{figure}

\begin{table}
\centering
\caption{Results of Augmented Dickey-Fuller (ADF) and KPSS Stationarity Tests for S4 0954+65 light curve.}
\label{tab:stationarity_blazar}
\begin{tabular}{lcc}
\hline\hline
\textbf{Statistic} & \textbf{ADF Test} & \textbf{KPSS Test} \\
\hline
Test Statistic & -3.971632 & 0.514357 \\
p-value & 0.001567 & 0.038433 \\
Lags Used & 4 & 18 \\
Number of Observations Used & 845 & -- \\
Critical Value (1\%) & -3.438112 & 0.739000 \\
Critical Value (5\%) & -2.864966 & 0.463000 \\
Critical Value (10\%) & -2.568595 & 0.347000 \\
\hline\hline
\end{tabular}
\end{table}
We therefore fix the differencing order to $d=1$ for the grid searches. The results of the grid search for both light curves are shown in Figure \ref{fig:quasar_heatmap} and Figure \ref{fig:blazar_heatmap}.
From the grid searches, ARIMA$(1,1,1)$ is picked as the model with highest log posterior probability ($\log{P_{\text{max}}}=-0.0243 \pm0.79$) for 3C 273 data and ARIMA$(1,1,4)$ ($\log{P_{\text{max}}}=-1.1964 \pm0.81$) for S4 0954+65. We therefore fit these models to the data with a high-resolution nested sampling run, and show the results in Figures \ref{fig:quasar_fit} and \ref{fig:blazar_fit}.
The residual time series for both 3C 273 and S4 0954+65 show no temporal structure. Figure \ref{fig:quasar_resid_hist} shows the histograms of the residual time series for 3C 273 and S4 0954+65.
The ACF and PACF plots of the light curves and residuals (see Figures \ref{fig:quasar_acf} and \ref{fig:blazar_acf}) show that the respective ARIMA model fits have successfully removed the correlation in the original light curves. The results of the Ljung-Box test on both residuals (see Tables \ref{tab:ljung_box_test_quasar} and \ref{tab:ljung_box_test_blazar}) also indicate no serial correlation. 
\begin{table}
\centering
\caption{Ljung--Box (LB) test p-values for residual autocorrelation of 3C 273 light curve.}
\label{tab:ljung_box_test_quasar}
\begin{tabular}{cc}
\hline\hline
\textbf{Lag} & \textbf{p-value} \\
\hline
1   & 0.858400 \\
2   & 0.443073 \\
3   & 0.632991 \\
4   & 0.677931 \\
5   & 0.743433 \\
6   & 0.778520 \\
7   & 0.764781 \\
8   & 0.755184 \\
9   & 0.709967 \\
10  & 0.579299 \\
\hline\hline
\end{tabular}
\end{table}
\begin{table}
\centering
\caption{Ljung--Box (LB) test p-values for residual autocorrelation of S4 0954+65 light curve.}
\label{tab:ljung_box_test_blazar}
\begin{tabular}{cc}
\hline\hline
\textbf{Lag} & \textbf{p-value} \\
\hline
1   & 0.161701 \\
2   & 0.293037 \\
3   & 0.252978 \\
4   & 0.361652 \\
5   & 0.474434 \\
6   & 0.565536 \\
7   & 0.674115 \\
8   & 0.630585 \\
9   & 0.568174 \\
10  & 0.649945 \\
\hline\hline
\end{tabular}
\end{table}
The resulting residual time series are therefore characteristic of a white noise sequence. Hence, we conclude that ARIMA$(1,1,1)$ and ARIMA$(1,1,4)$ have modelled the light curves of 3C 273 and S4 0954+65, respectively.
\section{Conclusions}\label{conclusions}
In this paper, we presented a vectorized ARIMA–Nested Sampling framework and validated it on simulated time series data, recovering true model orders and parameters accurately. The framework was also applied to sunspot counts and to light curve data from Kepler and TESS. It successfully characterised the stochastic variability in these light curves and produced well-calibrated forecasts for the sunspot counts. We note that future improvements to the multi-step forecasts could be made by adopting a rolling window, which is a more accurate (although computationally expensive in our case) way of implementing ARIMA forecasting \citep{rolling_window_forecasts}. Additionally, the forecasting can be made more robust within a Bayesian framework using a short time-step rolling forecast and consequently treating the obtained posterior distributions as priors for the next forecast iteration.

The dominant computational bottleneck was the rejection sampling of valid ARMA coefficient vectors ($\phi_a$ and $\theta_m$) for higher orders ($p+q>8$) using the process discussed in Section \ref{methodology}. If very strong autocorrelation in the time series is not expected \emph{a priori}, then the ``rejection sampling" process can be sped up by regularizing and tightening the priors using $\sigma''<1$ (Table \ref{tab:prior_table}). This is quantified in Appendix \ref{appendixb} (Table \ref{tab:benchmark}), which reports wall-clock runtimes for grid searches of increasing size and for different prior scales. A more powerful approach is offered by reparametrization of the ARMA$(p,q)$ coefficients to a set of $p+q$ partial autocorrelation coefficients \citep{BARNDORFFNIELSEN1973408}. These coefficients are naturally constrained between $-1$ and $1$, and therefore a simple uniform prior, bounded between these two values, can be used. The vectorized implementation also offers a significant speed-up through parallelization of the nested sampling controlled by the $n_\mathrm{delete}$ factor. These computation times can be further improved through GPU-acceleration which our framework supports.

We note that, while the framework enables joint selection across all three ARIMA orders $p$, $d$ and $q$, the real-data applications in this paper fixed $d$ \emph{a priori} using classical stationarity tests (ADF and KPSS) and performed grid searches over $p$ and $q$ only. Full three-dimensional grid searches including $d$ are demonstrated on simulated data (Section \ref{results}, Table \ref{logp_d_table}) and are entirely supported by the framework; we leave their application to real astronomical datasets with ambiguous stationarity as a direction for future work.

Although limited to classic ARIMA models, the framework presented here can be extended to its hybrid variants such as Seasonal ARIMA (SARIMA), Seasonal ARIMA with exogenous variables (SARIMAX), Continuous ARIMA (CARIMA) and Autoregressive Fractionally Integrated Moving Average (ARFIMA) models \citep{10.3389/fphy.2018.00080}. These extensions are efficient in modelling both periodic and stochastic variability exhibited in many astronomical time series, such as those of transiting exoplanets, eclipsing binaries or Active Galactic Nuclei (AGN).
\par
Finally, the investigation of potential physical interpretations behind the use of autoregressive modelling in time domain astronomy forms another compelling direction of future work. This question could be explored by applying our model selection procedure to different time-varying astrophysical systems. If systems of a particular class tend to favour similar model structures, then the ARIMA model may be representative of an underlying physical process rather than being a mere statistical description of the data.

\section*{Acknowledgements}
This material is based upon work supported by the Google Cloud research credits program, with the award GCP397499138. The authors were supported by the research environment and infrastructure of the Handley Lab at the University of Cambridge. AN was funded through the Cambridge Mathematics Placement (CMP) programme and the Institute of Astronomy summer research programme at the University of Cambridge. WH was supported by a Royal Society University Research Fellowship.

\section*{Data Availability}
The sunspots time series data used in this work were obtained from the \emph{sunspots} dataset provided in the \texttt{statsmodels} library (credit: SIDC, RWC Belgium, World Data Center for the Sunspot Index, Royal Observatory of Belgium, 1700-2008). The Kepler and TESS photometric light curves were obtained and processed using the \texttt{lightkurve} \citep{2018ascl.soft12013L} Python package. All the data used in this analysis, including the relevant nested sampling chains, are available at \citep{naik_2025_17771974}. We also include a python notebook to generate the plots and figures  presented in this paper.
\bibliographystyle{mnras}
\bibliography{bibliography}
\appendix
\section{Comparison of Forecasting Performance}\label{sunspots_comparison}
\begin{figure*}
    \centering
    \includegraphics{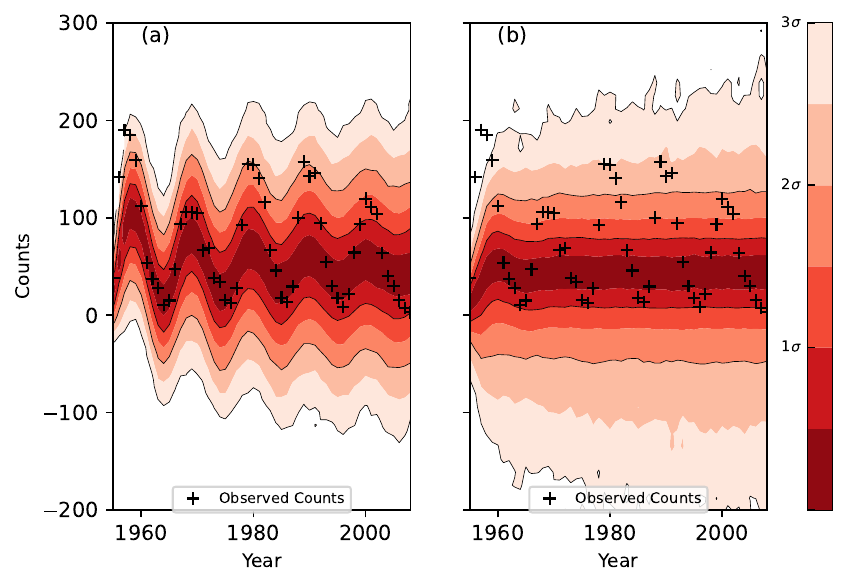}
    \caption{Comparison of posterior predictive forecasts for two ARIMA models applied to the yearly sunspot number time series. Panels (a) and (b) show the posterior predictive distributions of the ARIMA$(9,0,1)$ and ARIMA$(3,0,3)$ models, respectively with shaded contours denoting the $1\sigma$,$2\sigma$ and $3\sigma$ credible regions of the predictive posterior distribution $P(\hat{y}_t|t,D_t)$.} 
    \label{fig:forecast_compare}
\end{figure*}

The forecasting performance of ARIMA$(9,0,1)$ and ARIMA$(3,0,3)$ are compared in Figure \ref{fig:forecast_compare}.
ARIMA$(9,0,1)$ achieves a markedly better overall forecasting performance compared to ARIMA$(3,0,3)$. It shows the expected growth of predictive uncertainty with time, which is characteristic of long-term ARIMA forecasting. ARIMA$(9,0,1)$ forecasts also yielded much lower values for the Root Mean Squared Error (RMSE) and Mean Absolute Error (MAE) (see Table \ref{tab:arima_metrics}), thereby confirming and validating the results of our model selection methodology.

\begin{table}
\centering
\caption{Predictive performance metrics for the ARIMA($9,0,1$) and ARIMA($3,0,3$) models. ELPD denotes the Approximate Log Predictive Density calculated using a Gaussian predictive distribution. Higher values indicate better probabilistic calibration.}
\begin{tabular}{lcc}
\hline\hline
\textbf{Metric} & \textbf{ARIMA($9,0,1$)}  & \textbf{ARIMA($3,0,3$)} \\
\hline
Root Mean Squared Error (RMSE) & 42.25 & 60.11 \\
Mean Absolute Error (MAE)  & 32.24 & 46.85 \\
Log Predictive Density (ELPD) & -281.99 & -317.52\\

\hline\hline
\end{tabular}
\label{tab:arima_metrics}
\end{table}

\section{Computational Performance}\label{appendixb}
Table \ref{tab:benchmark} reports wall-clock runtimes for ARIMA grid searches of increasing size, evaluated on a synthetic time series of $N=1000$ data points. We use a standard MacBook Air (M2 chip, 16 GB RAM) for benchmarking. All runs used $n_\mathrm{live} = 100$ and $n_\mathrm{delete}=50$. 

\begin{table}
\centering
\small
\caption{Wall-clock runtimes (seconds) for ARIMA grid searches of 
increasing size on a synthetic time series of $N = 1000$ data points, 
evaluated on a standard consumer laptop (Apple M2 MacBook Air, 16 GB RAM). 
All runs used $n_\mathrm{live} = 100$ and $n_\mathrm{delete} = 50$. The timings are evaluated for different prior scales $\sigma''$ of the ARMA coefficients (Table \ref{tab:prior_table}).}
\label{tab:benchmark}
\begin{tabular}{cccc}
\hline
& \multicolumn{3}{c}{Total Time (s)} \\
\cline{2-4}
Grid & $\sigma''=1$ & $\sigma''=0.8$ & $\sigma''=0.6$ \\
\hline
$2\times2$ & 71.12 & 72.57 & 70.79 \\
$5\times5$ & 1046.20 & 1036.95 & 1027.45 \\
$7\times7$ & 3473.27 & 3318.80 & 3276.66\\
\hline
\end{tabular}
\end{table}
\bsp
\label{lastpage}
\end{document}